\documentclass[twocolumns]{aa}
\usepackage{natbib}
\bibpunct{(}{)}{;}{a}{}{,} 
\usepackage[varg]{txfonts}
\usepackage{graphicx}
\usepackage{gensymb}
\usepackage[utf8]{inputenc}

\graphicspath{{./figures/}}
\usepackage{caption}
\usepackage{subcaption}

\usepackage[switch]{lineno}

\begin{document}

\title{Radio timing constraints on the mass of the binary pulsar PSR~J1528$-$3146}

\author{A.~Berthereau\inst{1,2}
\and L.~Guillemot\inst{1,2}
\and P.~C.~C.~Freire\inst{3}
\and M.~Kramer\inst{3}
\and V.~Venkatraman~Krishnan\inst{3}
\and I.~Cognard\inst{1,2}
\and G.~Theureau\inst{1,2,4}
\and M.~Bailes\inst{5,6}
\and M.~C.~i~Bernadich\inst{3}
\and M.~E.~Lower\inst{7}
}

\institute{
Laboratoire de Physique et Chimie de l'Environnement et de l'Espace, Universit\'e d'Orl\'eans / CNRS, 45071 Orl\'eans Cedex 02, France\\
\email{anais.berthereau@cnrs-orleans.fr} 
\and
Observatoire Radioastronomique de Nan\c{c}ay, Observatoire de Paris, Universit\'e PSL, Université d'Orl\'eans, CNRS, 18330 Nan\c{c}ay, France
\and
Max-Planck-Institut f\"ur Radioastronomie, Auf dem H\"ugel 69, D-53121, Bonn, Germany
\and
LUTH, Observatoire de Paris, Universit\'e PSL, Universit\'e Paris Cit\'e, CNRS, 92195 Meudon, France
\and
Centre for Astrophysics and Supercomputing, Swinburne University of Technology, PO Box 218, Hawthorn, Vic, 3122, Australia 
\and
ARC Centre of Excellence for Gravitational Wave Discovery, OzGrav 
\and 
Australia Telescope National Facility, CSIRO, Space and Astronomy, PO Box 76, Epping, NSW 1710, Australia
}

\authorrunning{A.~Berthereau} 

\abstract
{PSR~J1528$-$3146 is a 60.8 ms pulsar orbiting a heavy white dwarf (WD) companion, with an orbital period of 3.18 d. The pulsar was discovered in the early 2000s in a survey at 1.4~GHz of intermediate Galactic latitudes conducted with the Parkes radio telescope. The initial timing analysis of PSR~J1528$-$3146 used data recorded from 2001 and 2004, and did not reveal any relativistic perturbations to the orbit of the pulsar or to the propagation of its pulses. Yet, with an orbital eccentricity of $\sim 0.0002$ and a large companion mass in the order of 1 M$_\odot$, this system was likely to exhibit measurable perturbations.}
{This work aimed at characterizing the pulsar's astrometric, spin and orbital parameters by analyzing timing measurements conducted at the Parkes, MeerKAT and Nan\c{c}ay radio telescopes over almost two decades. The measurement of post-Keplerian perturbations to the pulsar's orbit can be used to constrain the masses of the two component stars of the binary, and in turn inform us on the history of the system.}
{We analyzed timing data from the Parkes, MeerKAT and Nan\c{c}ay radio telescopes collected over $\sim$16 yrs, obtaining a precise rotation ephemeris for PSR~J1528$-$3146. A Bayesian analysis of the timing data was carried out to constrain the masses of the two components and the orientation of the orbit. We further analyzed the polarization properties of the pulsar, in order to constrain the orientations of the magnetic axis and of the line-of-sight with respect to the spin axis. }
{We measured a significant rate of advance of periastron for the first time, and put constraints on the Shapiro delay in the system and on the rate of change of the projected semi-major axis of the pulsar's orbit. The Bayesian analysis yielded measurements for the pulsar and companion masses of respectively $M_p = 1.61_{-0.13}^{+0.14}$ M$_\odot$ and $M_c = 1.33_{-0.07}^{+0.08}$ M$_\odot$ (68\% C.L.), confirming that the companion is indeed massive. This companion mass as well as other characteristics of PSR~J1528$-$3146 make this pulsar very similar to PSR~J2222$-$0137, a 32.8 ms pulsar orbiting a WD whose heavy mass ($\sim 1.32$ M$_\odot$) was unique among pulsar-WD systems until now. Our measurements therefore suggest common evolutionary scenarios for PSRs J1528$-$3146 and J2222$-$0137.}
{}


\maketitle
\let\oldpageref\pageref
\renewcommand{\pageref}{\oldpageref*}
\newcommand{\nrt}{Nan\c{c}ay }
\newcommand{\psr}{PSR~J1528$-$3146 }
\hyphenchar\font=-1


\section{Introduction}

\label{Introduction}
Pulsars are highly magnetized, rapidly-rotating neutron stars that emit periodic trains of radio pulses. The pulsar population can be divided into two main groups: ``normal'' pulsars and ``millisecond'' pulsars (MSPs); the latter being pulsars with small rotational periods ($P \lesssim 30$ ms), and very small period increases ($\dot P \lesssim 10^{-17}$ s s$^{-1}$). In most cases MSPs are found to be members of binary systems, and are thought to have been ``recycled'' by the accretion of matter and transfer of angular momentum from their binary companion, spinning up their rotation to millisecond periods \citep{BKK1974,alpar1982}. In some cases, this accretion process can stop before the pulsar gets fully recycled, leading to so-called ``mildly-recycled'' pulsars with rotational periods between $20$ and $100$ ms. This process happens mostly if the companion stars are more massive: such stars evolve more rapidly and therefore any accretion episodes will generally be much shorter.

\psr is a member of the latter sub-class of mildly-recycled pulsars. This $P \sim 60.8$ ms pulsar was discovered in a survey conducted at 1.4~GHz with the Parkes radio telescope (also known as 'Murriyang') in Australia, as part of an extension of the Swinburne Intermediate Latitude survey, which discovered a total of 26 pulsars \citep{Jacoby_discovery}. Timing measurements conducted from 2001 to 2004 revealed that the pulsar is in a binary system with an orbital period $P_b \sim 3.18$ days, a projected semi-major axis $x \equiv a \sin i /c \sim 11.4\,$s (where $a$ is the semi-major axis of the pulsar's orbit, $i$ is the orbital inclination and $c$ is the speed of light) and an orbital eccentricity $e \sim 0.0002$ which is large for systems with comparable orbital periods \citep[e.g.,][]{Berezina2017}. The orbital period and semi-major axis imply a relatively massive companion: assuming a pulsar mass of 1.4 M$_\odot$, the mass function of 0.159 M$_\odot$ indicates that the minimum companion mass (calculated for an edge-on orbit, \textit{i.e.}, with an inclination of $90\degree$) is $\sim 0.94$ M$_\odot$.  With such properties the system can be classified as an intermediate-mass binary pulsar \citep[IMBP, see e.g.][]{Camilo1996,Tauris2012}; the large mass of the companion suggests that it must be a CO or ONeMg white dwarf (WD). Optical observations of the field around \psr with the 6.5 m Baade telescope at the Magellan Observatory led to the detection of a potential counterpart, with an R-band magnitude of about $24.2$ and a B-band magnitude of about 24.5\citep{Jacoby2006}. If the counterpart is indeed associated with the J1528$-$3146 system, then the optical emission very likely originates from the WD.

The initial timing study by \citet{Jacoby_discovery} over approximately three years did not reveal any relativistic perturbations to the pulsar's orbit, such as for instance the rate of periastron advance. Nevertheless, the non-negligible orbital eccentricity and large companion mass make it a promising system for measuring such relativistic perturbations, that can be described in pulsar timing analyses by so-called ``Post-Keplerian'' (PK) parameters \citep[e.g.,][]{DT1992}. 

Two decades after the discovery of PSR~J1528$-$3146, the accumulation of high-quality pulsar timing data by the Nan\c{c}ay radio telescope (NRT) in France, and the advent of highly-sensitive new generation instruments such as the MeerKAT 64-dish array in South Africa \citep{Jonas2009} and the Ultra-Wideband Low receiver at the Parkes telescope \citep[UWL, see][]{UWL} motivated a new study of this pulsar and of its timing properties. In particular, a new timing study of \psr should enable more precise measurements of its parameters, and could yield first detections of PK parameters in this system and possibly measurements of the masses of the pulsar and of its companion.

In this article we report on the results from the analysis of a $\sim$16 yr timing data set on \psr with the Parkes, NRT and the MeerKAT radio telescopes.
The article is organized as follows: in Section~\ref{Data} we describe the details of the observations of \psr and the extraction of pulsar timing data from these observations. In Section~\ref{Polarization} we present an analysis of the linear polarization properties of the pulsar, and the implications of the results in terms of emission geometry. In Section~\ref{Timing} we present the main timing results from our study and the constraints on the mass of the pulsar and of its companion. We finally summarize our results and discuss on the main implications from this study in Section~\ref{Discussion}. 


%

\section{Data}
\label{Data}

\subsection{Observations}

\begin{figure*}[ht]
\begin{center}
\includegraphics[width=0.48\textwidth]{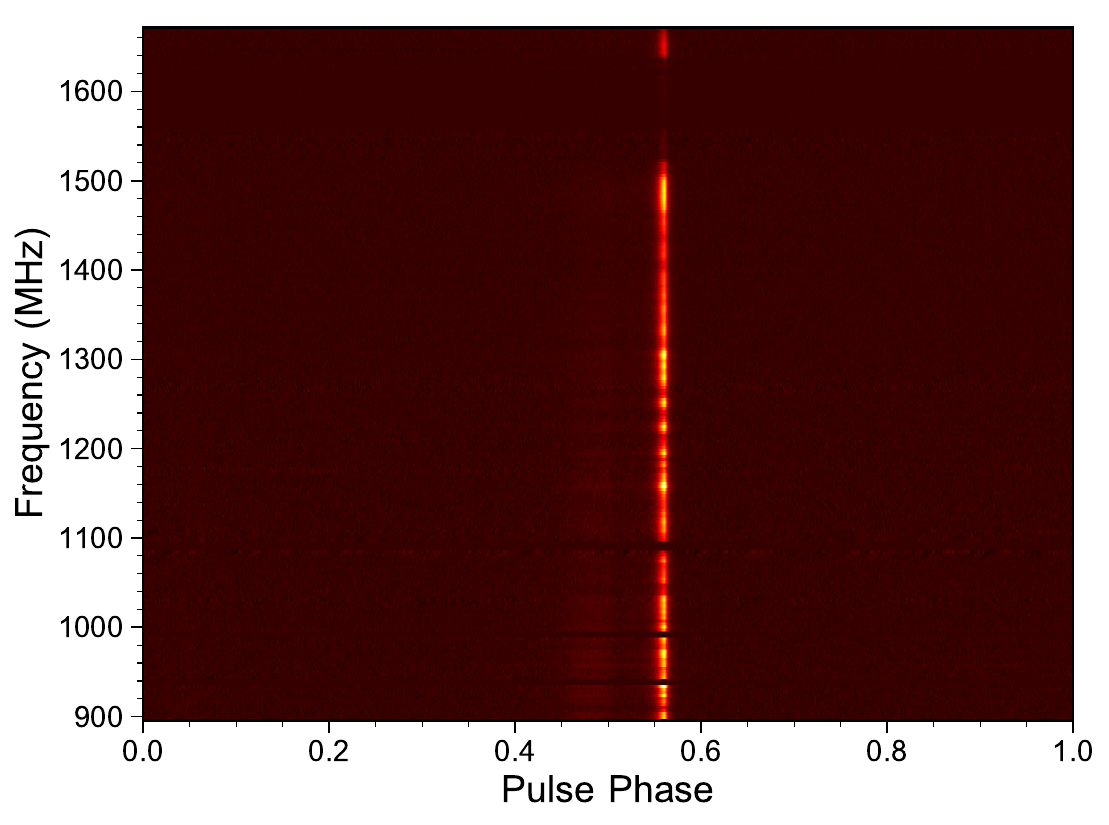}
\includegraphics[width=0.48\textwidth]{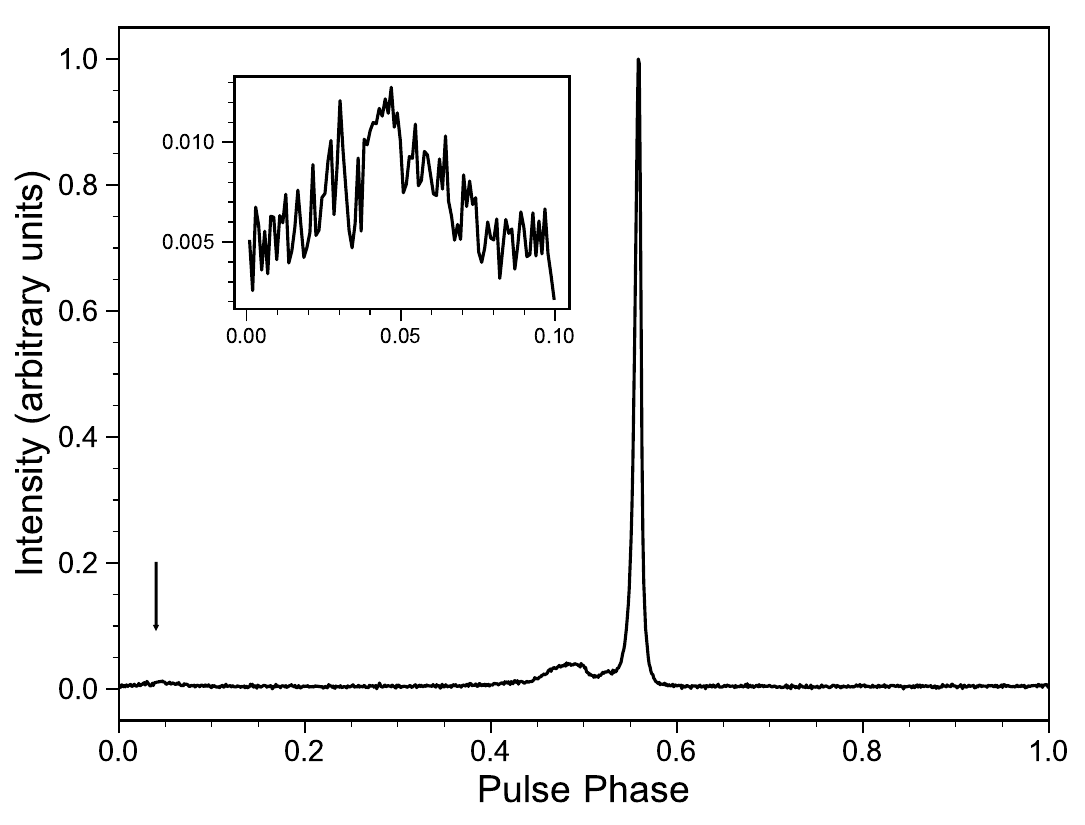}
\caption{MeerKAT pulse profiles at 1284~MHz for PSR~J1528$-$3146, from a $\sim 4$-hr observation at MJD~58829, corresponding to the highest S/N observation in our dataset. \textit{Left:} total intensity as a function of observing frequency and rotational phase. \textit{Right}: integrated pulse profile, with the main radio pulse centered at rotational phase $\sim 0.54$, and a previously-unknown weak radio component at phase $\sim 0.03$, indicated with an arrow. The inset at the top left shows a zoom-in of the pulse profile around the weak radio component.}
\label{fig:meerkatprofile}
\end{center}
\end{figure*}

The NRT dataset consisted in observations made with the Berkeley-Orl\'eans-Nan\c{c}ay (BON) backend until August 2011, and the NUPPI (a version of the Green Bank Ultimate Pulsar Processing Instrument, GUPPI, made for the NRT) backend afterwards.

The BON observations of \psr started in November 2006, and were done at a central frequency of 1398~MHz. These observations were coherently dedispersed and were done over a frequency bandwidth of 64~MHz until July 2008, and over a bandwidth of 128~MHz after that date.

The NUPPI instrument, which became the primary pulsar timing instrument at \nrt after August 2011, increased the bandwidth of NRT pulsar observations to 512~MHz, also coherently dedispersed. NUPPI observations were conducted at a central frequency of 1484~MHz. In this study we considered NUPPI data recorded until May 2022. Although pulsar observations with the BON backend were maintained for another few years after 2011 for testing purposes, we discarded BON observations made simultaneously with NUPPI observations, as they represented duplicated astrophysical information.

The Parkes radio telescope dataset included the timing data presented in \citet{Jacoby_discovery}, recorded between September 2001 and September 2004, and more recent data taken with the UWL recorded from October 2019 to January 2021. The former observations have been conducted with the 21-cm multibeam receiver \citep{Staveley1996} at a central frequency of 1390 MHz, using the $2 \times 0.5 \times 512$
~MHz analogue filterbank backend (AFB) and hence were incoherently dedispersed. Observations with the UWL receiver were carried out using the \textsc{MEDUSA} backend, that performed coherent dedispersion over a much larger bandwidth of 3328~MHz, centered at 2368~MHz.

The MeerKAT dataset we used for the study comprised observations made between March 2019 and April 2021, as part of the RelBin program of the MeerTime large science project. The RelBin program aims at improving the measurement and detecting new relativistic effects in known pulsars, increasing the number of pulsars with mass measurements, and testing gravity theories, with the excellent timing capabilities of MeerKAT \citep{Kramer-Relbin}. Pulsar observing set-ups, verifications and early science results with MeerKAT are presented in \citet{Bailes2020}. Observations were calibrated in polarization as described in \citet{Serylak2021}. MeerKAT observations were conducted at a central frequency of 1284~MHz, and covered a frequency bandwidth of about 800~MHz. While the bulk of the observations had a typical duration of about 30~minutes, an intensive orbital campaign was conducted around 2019 December 12 (MJD~58829). On that day, the system was observed for $\sim$ 4~hrs to cover a wide range of orbital phases centered on superior conjunction in order to significantly increase the sensitivity of our subsequent timing analyses to a faint Shapiro delay \citep{Shapiro1964} signal that had already been detected in Nan\c{c}ay data. Additionally, three other 30~min observations of the pulsar were conducted within three days centered on MJD 58829, in order to sample orbital phase ranges away from superior conjunction.

In Figure~\ref{fig:meerkatprofile} we show the MeerKAT pulse profile for \psr from the latter observation, which represented the highest S/N observation in our total dataset. The main radio component in the pulse profile is at rotational phase $\sim 0.54$ in this graph. As can be seen from Figure~\ref{fig:meerkatprofile}, the unprecedented sensitivity of MeerKAT observations has enabled us to detect a previously unseen weak component separated from the main pulse by approximately half a rotation. The presence of this interpulse component provides useful information about the pulsar's emission geometry, as will be discussed further in Section~\ref{Polarization}.

\subsection{Data analysis}

\begin{table*}
\caption{Summary of observation parameters and TOA properties for each of the datasets considered in the timing study.}
\label{tab:ToAs}
\centering
\begin{tabular}{lccccc}
\hline
\hline
Dataset & Parkes AFB & Parkes UWL & MeerKAT & Nan\c{c}ay BON & Nan\c{c}ay NUPPI \\
\hline
Start of observations (MJD) & 52179 & 58759 & 58557 & 54065 & 55848 \\
End of observations (MJD) & 53262 & 59216 & 59319 & 55819 & 59725 \\
Observing frequency (MHz) & 1398 & 2368 & 1284 & 1398 & 1484 \\
Frequency bandwidth (MHz) & 256 & 3328 & 776/856\tablefootmark{c} & 64/128\tablefootmark{d} & 512 \\
Number of TOAs & 169 & 53 & 936 & 866 & 11454 \\
Bandwidth per TOA (MHz) & 128 & 3328\tablefootmark{b} & 97/107\tablefootmark{c} & 32/64\tablefootmark{d} & 64 \\
Time per TOA (s) & ---\tablefootmark{a} & 600 & 600 & 600 & 600 \\
Median TOA uncertainty ($\mu$s) & 19.3 & 3.2 & 3.5 & 30.4 & 29.0\\
Weighted residual rms ($\mu$s) & 12.8 & 4.3 & 3.0 & 16.3 & 16.3 \\
EFAC & 1.13 & 1.69 & 1.22 & 0.98 & 0.98 \\
\hline
\end{tabular}
\tablefoot{
\tablefoottext{a}{These observations were scrunched in time to single sub-integrations. See \citet{Jacoby_discovery} for details on the corresponding observations.}
\tablefoottext{b}{For this dataset the TOAs were generated using the wide-band template matching technique implemented in \texttt{PulsePortraiture} \citep[see][]{Pennucci2014}.}
\tablefoottext{c}{The bulk of the MeerKAT observations covered a total of 856~MHz of bandwidth. For these observations we grouped the channels in eight sub-bands of 107~MHz each. Other observations covered a slightly shorter bandwidth of 776~MHz, for these we also grouped the channels in eight sub-bands of 97~MHz.}
\tablefoottext{d}{The original frequency bandwidth for the BON backend of 64~MHz increased to 128~MHz in July 2008. For observations taken prior to July 2008, the bandwidth of 64~MHz was split into 32~MHz sub-bands, for the other observations the bandwidth of 128~MHz was split into 64~MHz sub-bands.}
}
\end{table*}

All data reduction steps for the datasets described above were done using the \textsc{PSRCHIVE} analysis library \citep{psrchive}. Observations were cleaned of radio frequency interference (RFI) and polarization-calibrated.

The typically $\sim 1$-hr long \nrt observations with the BON and NUPPI backends were split into sub-integrations of 10-min each. For the BON dataset, these sub-integrations were split into two frequency sub-bands of 32~MHz each for the data recorded before July 2008, and two sub-bands of 64 MHz for the data taken after the bandwidth increase. Sub-integrations for the NUPPI data were divided into eight frequency sub-bands of 64~MHz each. For the BON and NUPPI data we combined 10 observations with the highest signal-to-noise ratios (S/N) that we then smoothed to produce noise-free standard profiles at 1.4~GHz with the same frequency resolution. Times of arrival (TOAs) to be used in the timing analysis (described in Section~\ref{Timing}) were derived from the \nrt data using the Matrix Template Matching (MTM) method of the \texttt{pat} routine of \textsc{PSRCHIVE} \citep{vanStraten2006}.

TOAs were extracted from the ``historic'' Parkes dataset by comparing observations with a high S/N template profile, as described in \citet{Jacoby_discovery}. Each observation was split into two frequency sub-bands of 128~MHz, so that two TOAs were extracted from each observation. On the other hand, the very large frequency bandwidth covered by Parkes observations with the UWL prompted us to consider a different approach for extracting TOAs. Instead of splitting the very large bandwidth into several sub-bands, which often resulted in sub-bands containing no clear detections at the highest frequencies, we used the wide-band template matching technique implemented in the \texttt{PulsePortraiture} library\footnote{\url{https://github.com/pennucci/PulsePortraiture}} \citep{Pennucci2014} and extracted one TOA for the entire bandwidth, every 10~min of observation.

Finally, the MeerKAT observations were divided into eight frequency sub-bands 97 MHz (resp. 107 MHz) for obsservations with a 759 MHz (resp. 856 MHz) bandwidth and the sub-integrations were grouped into segments of 10-min. For each resulting element of time and frequency we extracted a TOA using the standard Fourier Domain Markov Chain Monte Carlo (FDM) method of \texttt{pat}, comparing the profiles with a smoothed version of the highest S/N MeerKAT observation with the same number of frequency sub-bands. A summary of the TOA dataset produced for the timing analysis is presented in Table~\ref{tab:ToAs}.

\section{Pulse Polarization Analysis}
\label{Polarization}

\begin{figure}[ht]
\begin{center}
\resizebox{\hsize}{!}{\includegraphics[width=0.95\columnwidth]{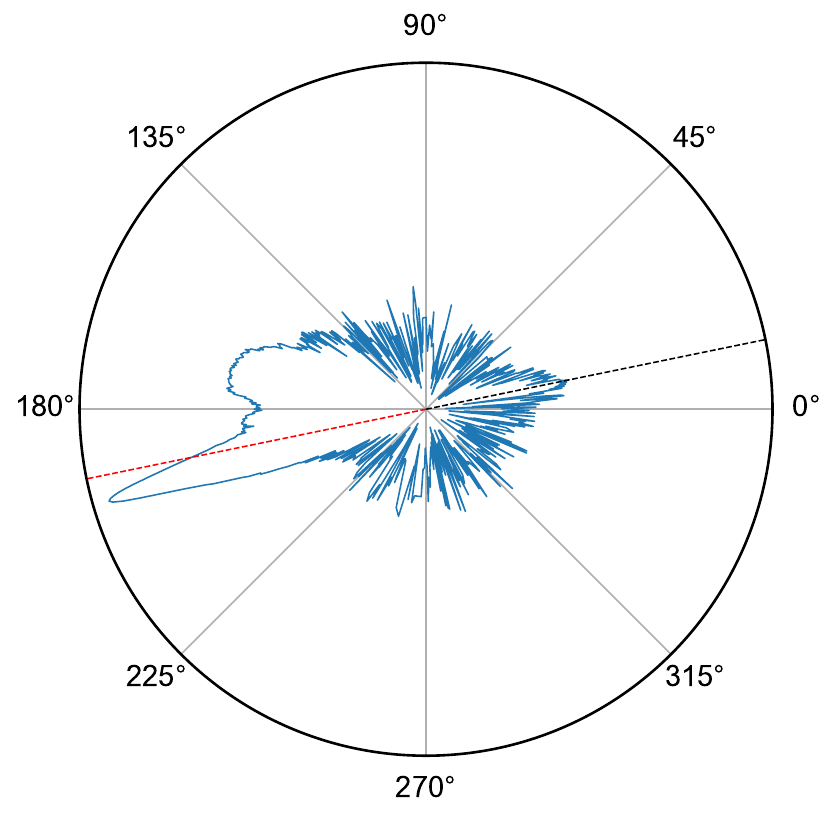}}
\caption{Polar representation of the MeerKAT pulse profile at 1284~MHz displayed in Figure~\ref{fig:meerkatprofile}, with radio flux densities shown in logarithmic scale. The black dashed line shows the position of the interpulse component, and the red dashed line shows the same position shifted by 180$^\circ$.}
\label{fig:meerkatprofile_polar}
\end{center}
\end{figure}

As mentioned in Section~\ref{Data}, the unprecedented sensitivity of MeerKAT radio observations of \psr have enabled us to detect a weak profile component well separated from the main emission pulse, for the first time. This newly-detected weak component can be seen in the polar representation of the pulse profile from the MeerKAT observation at MJD~58829 shown in Figure~\ref{fig:meerkatprofile_polar}. The dashed lines show the position of this profile component, and that of the same position shifted by half a rotation. The phase location was determined from a fit of the pulse profile displayed in Figure~\ref{fig:meerkatprofile} with a lorentzian line for the highest peak and two gaussian lines for the other two components, on top of a constant baseline. Denoting $\phi_\mathrm{M,1}$ and $\phi_\mathrm{M,2}$ as the phases of the two components of the main pulse, and $\phi_\mathrm{IP}$ that of the weak component, we found $\phi_\mathrm{M,1} = 0.469$(2), $\phi_\mathrm{M,2} = 0.54410$(2), and $\phi_\mathrm{IP} = 0.033$(8) from the fit of the pulse profile. The new component is therefore separated from the main emission cone by almost exactly half a rotation, and can thus be considered as interpulse emission, likely originating from the magnetic pole opposite to that responsible for the main emission pulse. The presence of this interpulse component indicates that the angle between the pulsar's spin axis and its magnetic axis, $\alpha$, must be close to 90$^\circ$.

Tighter constraints on the value of $\alpha$ and that of $\zeta$, the angle between the pulsar's spin axis and the observer's line-of-sight, may be placed by fitting the variation of the polarization position angle (PPA) $\Psi$ of the linearly polarized component of the pulsar as a function of rotational phase, with the Rotating Vector Model \citep[RVM,][]{RVM}. The position angles $\Psi$ are related to the $Q$ and $U$ Stokes parameters of the pulsar's emission as $\Psi = \frac{1}{2} \arctan{\left(\frac{U}{Q}\right)}$. In the following we use the modified form introduced in \citet{JK2019}:

\begin{equation}
\Psi = \Psi_0 + \arctan{\dfrac{\sin \alpha \sin \left(\phi - \phi_0 - \Delta\right)}{\sin \zeta \cos \alpha - \cos \zeta \sin \alpha \cos \left(\phi - \phi_0 - \Delta\right)}}.
\end{equation}
In the above expression, $\phi_0$ denotes the pulse phase for which $\Psi = \Psi_0$ (assuming $\Delta = 0$), and the term $\Delta$ can take account of a potential difference in emission heights between the two magnetic poles of the pulsar. We used PPA values as a function of rotational phase extracted from the 1284~MHz MeerKAT observation at MJD~58829. We first formed a time-averaged and dedispersed version of the observation, keeping the initial frequency resolution of 928 channels over the total bandwidth of 776~MHz. We used the ``Bayesian Lomb-Scargle Periodogram'' (BGLSP) method of \texttt{RMcalc}\footnote{\url{https://gitlab.mpifr-bonn.mpg.de/nporayko/RMcalc}} to fit for the pulsar's rotation measure, RM, which quantifies the amount of Faraday rotation occurring along the line-of-sight caused by the presence of magnetic field in the interstellar medium. Denoting $\lambda$ as the radio wavelength, we have: 

\begin{equation}
\Delta \Psi_\mathrm{PPA} = \mathrm{RM} \times \lambda^2.
\end{equation}

\begin{figure*}[ht]
\begin{center}
\includegraphics[width=0.95\textwidth]{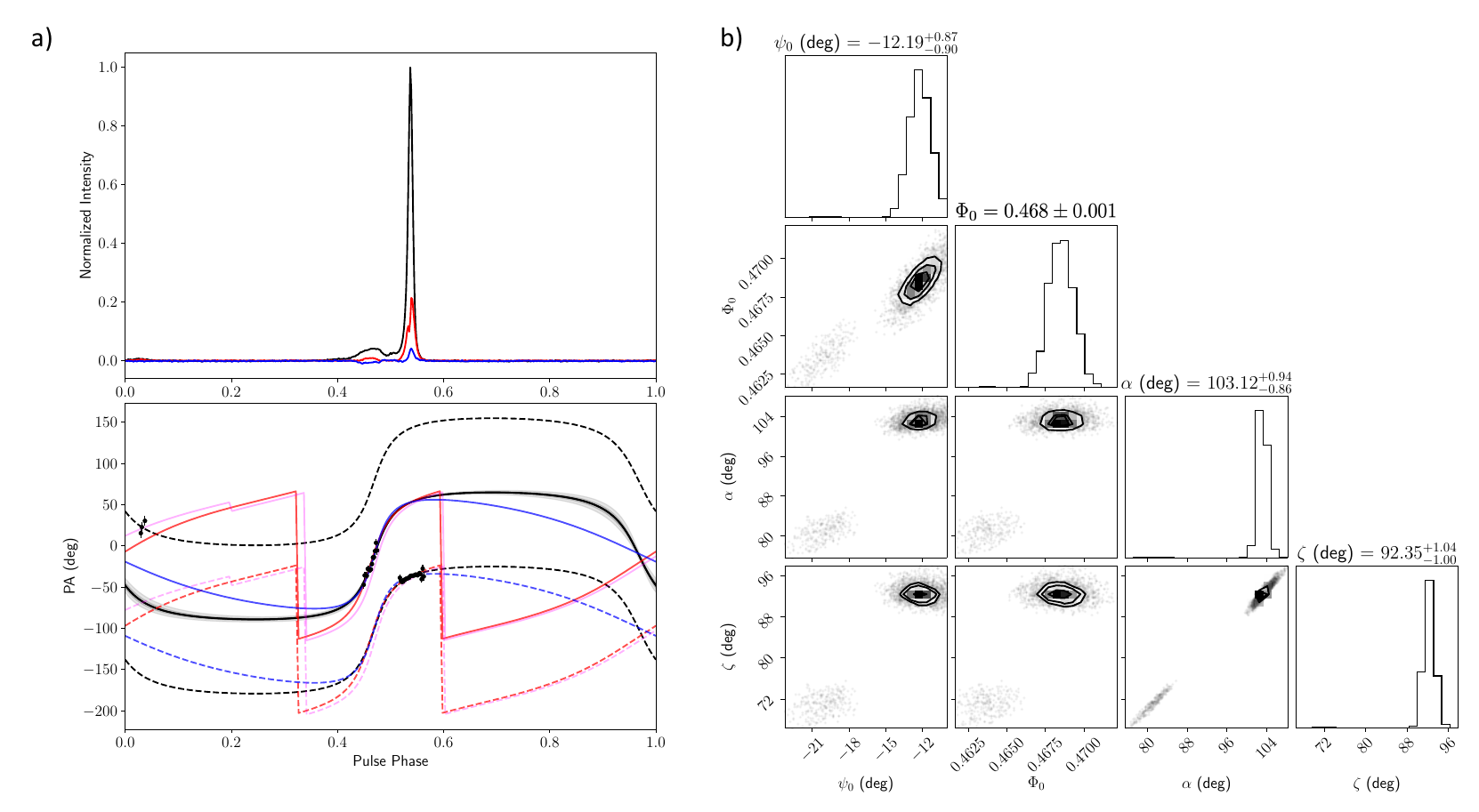}
\caption{\textit{Left:} the top panel shows the polarization profile of \psr{} as observed at 1284~MHz. The black
line represents total power, the red shows linearly and the blue line circularly polarized intensity, respectively.
The bottom panel displays the derived PPA values and a number of RVM fits. The black lines represents the resulting RVM for an unconstrained fit covering the whole parameter space. The dashed lines indicate vertical
shifts of the RVM by $\pm90^\circ$ to represent possible orthogonal polarised modes. The red and blue curves are RVM
fits for a restricted range of $\zeta$, forcing it to be consistent with values of the orbital inclination angle as derived from timing in this paper. The magenta line allows for a possible shift in emission height between the main
pulse and the interpulse. See text for details. \textit{Right:} corner plot of
the posterior distributions of the joint RVM model parameters, with the off-diagonal
elements representing the correlations between parameters, and the diagonal elements denoting the marginalised histograms. The posteriors shown here are derived for the unconstrained fit leading to the black line on the left.}
\label{fig:polfig}
\end{center}
\end{figure*}

Our BGLSP analysis yielded an RM value of $-19(1)$ rad~m$^{-1}$, consistent with the value published in \citet{Kramer-Relbin}. We finally corrected the values of $\Psi$ across the frequency bandwidth of the MeerKAT observation for the above-described Faraday rotation term, and formed a frequency-averaged pulse profile for \psr with the four Stokes parameters. Using this profile, we computed
the linearly and circularly polarised intensity as shown as red and blue lines in Figure~\ref{fig:polfig}a
(top panel), together with the total intensity as a black line. We then performed several RVM fits to the derived 
PPA data shown in the bottom panel of Figure~\ref{fig:polfig}a. The fits were obtained using the UltraNest sampler by \citet{ultranest}, using different priors for the angles in different runs. In each case, we used a standard
expression, $\ln({\cal L}) = -\chi^2/2$, as a likelihood function. In the first fit, we assumed our random
$\alpha$ and $\zeta$ angles to form spin and magnetic field vectors that are distributed uniformly over a unit-sphere,
while $\Delta=0$. The prior for $\Psi_0$ was a uniform distribution over $[-90^\circ,90^\circ]$ and, similarly,
$\phi_0$ had a uniform prior over $[0,1]$.
A corner plot of the obtained posterior distributions of the joint RVM model parameters, with the off-diagonal
elements representing the correlations between parameters and the
diagonal elements denoting the marginalised histograms is shown in 
Figure~\ref{fig:polfig}b. The PPA values of the discovered interpulse component help significantly to
constrain the derived parameter range. In addition to a solution at 
$(\alpha, \zeta) = (\sim103^\circ$, $\sim92^\circ$), a second solution, less favoured by the log-likelihood statistic, is seen for $(\alpha, \zeta) = (\sim80^\circ$, $\sim72^\circ$). Either case suggests an orthogonal rotator, as expected
if an interpulse is visble. The solid black line in the lower panel of Figure~\ref{fig:polfig}a is the 
derived RVM
corresponding to the preferred solution (with the gray band indicating 68\% uncertainties). The dashed black
lines show the RVM shifted by $\pm90^\circ$. If this fit is accurate, it indeed suggests that the main pulse
and interpulse emit in different orthogonal emission states, as found quite frequently in interpulse-pulsars
\citep[see e.g.][]{JK2019}.

As we will see in Section \ref{sec:Bayesian}, the Bayesian analysis of the timing data suggests 
an orbital inclination angle of either $i\sim 57^\circ$ or  $i\sim 123^\circ$. For a fully recycled pulsar, one
can expect the spin axis of the pulsar to be aligned with the orbital angular momentum vector, so
that $i \approx \zeta$, and also $\Psi_0 \approx \Omega$, where $\Omega$ is the longitude of the ascending node (see  Section \ref{sec:Bayesian}).
The usage of polarization data and RVM fits to determine the orbital 
configuration has been discussed recently, for instance, by \citet{Kramer-Relbin} or \citet{Guo2021}. 
Clearly, the above $\zeta$-values are not consistent with the results from timing. In a second and third
fit, we therefore applied uniform priors for $\zeta$ restricted to a $6^\circ$ interval centred on the respective
inclination angle values. The red curves are the results for $\zeta \sim 57^\circ$
($\alpha=69(1)^\circ$, $\zeta = 59(1)^\circ$, $\Psi_0=-23(1)^\circ$), while the blue lines
are for $\zeta \sim 123^\circ$ 
($\alpha=128(1)^\circ$, $\zeta = 120(1)^\circ$, $\Psi_0=-10(1)^\circ$), respectively. 
The latter case, results in an "outer line of sight", whereas
the former is an "inner line of sight" \citep[see e.g.][]{handbook}. It is clear that 
the red curve, $\zeta \sim 57^\circ$,
is a better fit to the data, but still far from perfect. 

One can improve the fit by allowing $\Delta \neq 0$. This is done in our last fit, the result of which is shown as 
the magenta line in the PPA panel, resulting in 
$\Delta=0.07(2)$ and a significant difference in emission height. 
($\alpha=67(2)^\circ$, $\zeta = 59(2)^\circ$, $\Psi_0=-25(1)^\circ$). Regardless of a non-zero $\Delta$ or not,
overall the solution for $\zeta\sim57^\circ$ describes the data much better than the solution with $i=123^\circ$.
It is interesting to study also the values for $\Omega$ derived from timing and listed in Table~\ref{tab:chi2params}.
Our value derived for an RVM with $\zeta$ close to $57^\circ$, $\Psi_0 = -23^\circ$ is consistent with a value
for $\Omega$ in the range derived for Solution 4 in Table~\ref{tab:chi2params}. 
Nevertheless, the overall quality of the fit when forcing
$\zeta$ to lie close to the derived orbital inclination angle is clearly limited. An unconstrained fit prefers a value much 
closer to $90^\circ$. 
As will be shown in Section~\ref{sec:Bayesian}, an orbital inclination value close to 90$^\circ$ is excluded by the timing analysis of PSR~J1528$-$3146.
Whether the applicability of the RVM is at fault here, or whether the assumption of aligned spin-vectors is wrong, are interesting questions to be discussed further.

\section{Timing Analysis}
\label{Timing}

The timing analysis was carried out by analyzing the topocentric TOA data presented in Section~\ref{Data}, using the \textsc{Tempo2} pulsar timing package \citep{TEMPO2}. The measured TOAs were first converted to Barycentric Coordinate Time (TCB), taking the known clock corrections for the different telescopes and using the JPL DE438 solar system ephemeris\footnote{\url{https://naif.jpl.nasa.gov/pub/naif/JUNO/kernels/spk/de438s.bsp.lbl}}. The orbital motion of the pulsar was described using the DDH orbital model, which uses the orthometric parameterization of the Shapiro delay \citep{Freire2010}. This parameterisation is well-suited for systems with low orbital inclinations (which is the case of PSR~J1528$-$3146, as will be shown later in this Section). The reason is that it describes the Shapiro delay using two parameters, the orthometric amplitude ($h_3$) and the orthometric ratio ($\varsigma$) that are, in such systems, much less correlated than the range ($r$) and shape ($s$) parameters of the DD orbital model \citep{Damour1985,Damour1986}. 

\begin{figure*}[ht]
\begin{center}
\includegraphics[width=0.95\textwidth]{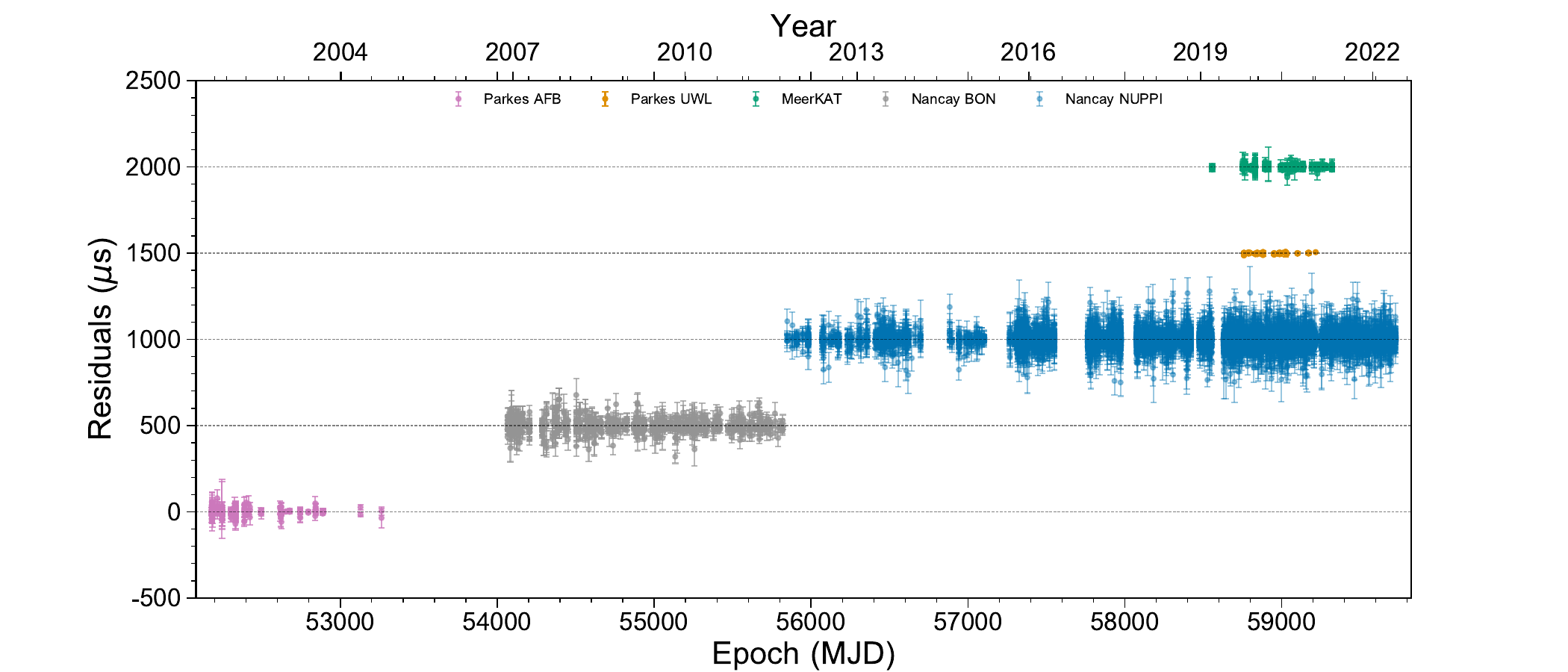}
\includegraphics[width=0.95\textwidth]{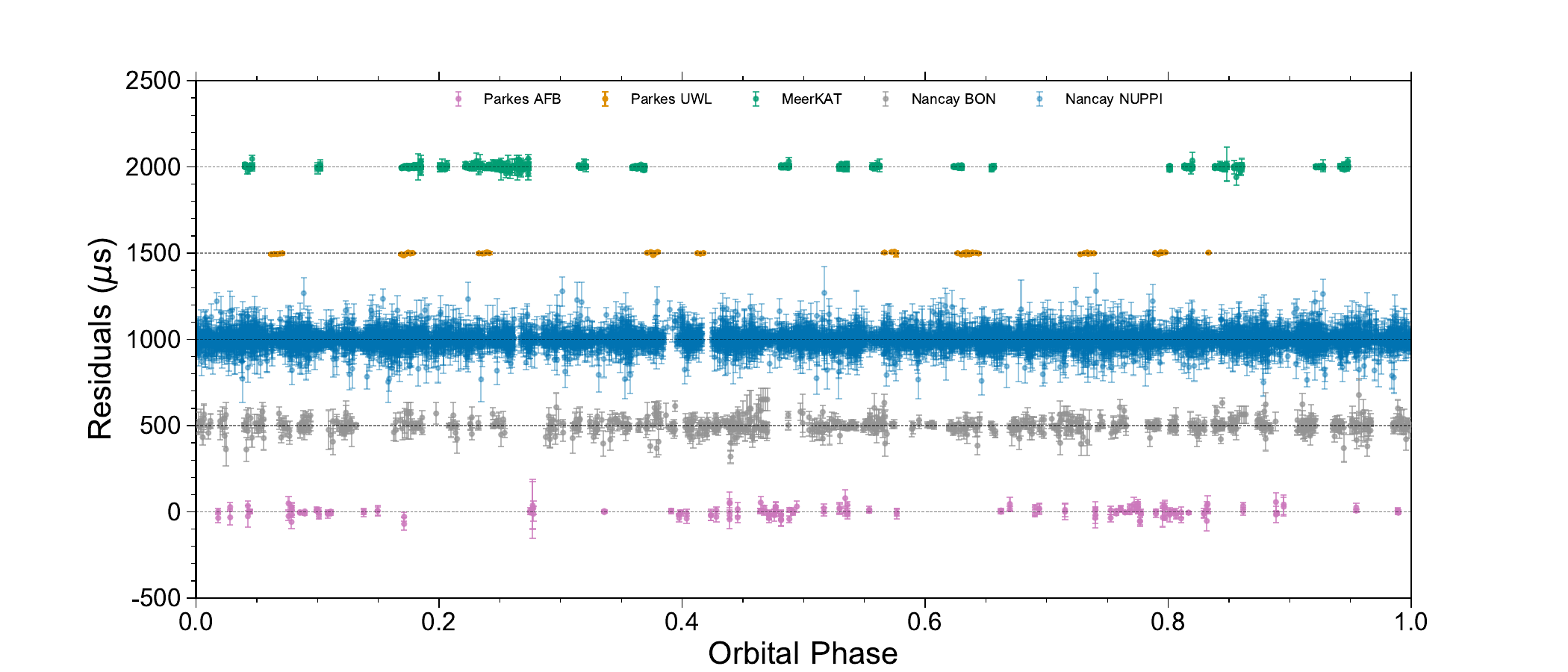}
\caption{Timing residuals as a function of time (top) and orbital phase (bottom), with the datasets considered in our analysis shifted by different amounts for ease of comparison. In the bottom plot, the orbital phase corresponding to superior conjunction (\textit{i.e.}, the moment of the orbit when the pulsar is on the opposite side of the companion) was shifted to occur at 0.25.}
\label{fig:toas_dates_ophases}
\end{center}
\end{figure*}

\begin{figure*}[ht]
\begin{center}
\includegraphics[width=0.95\textwidth]{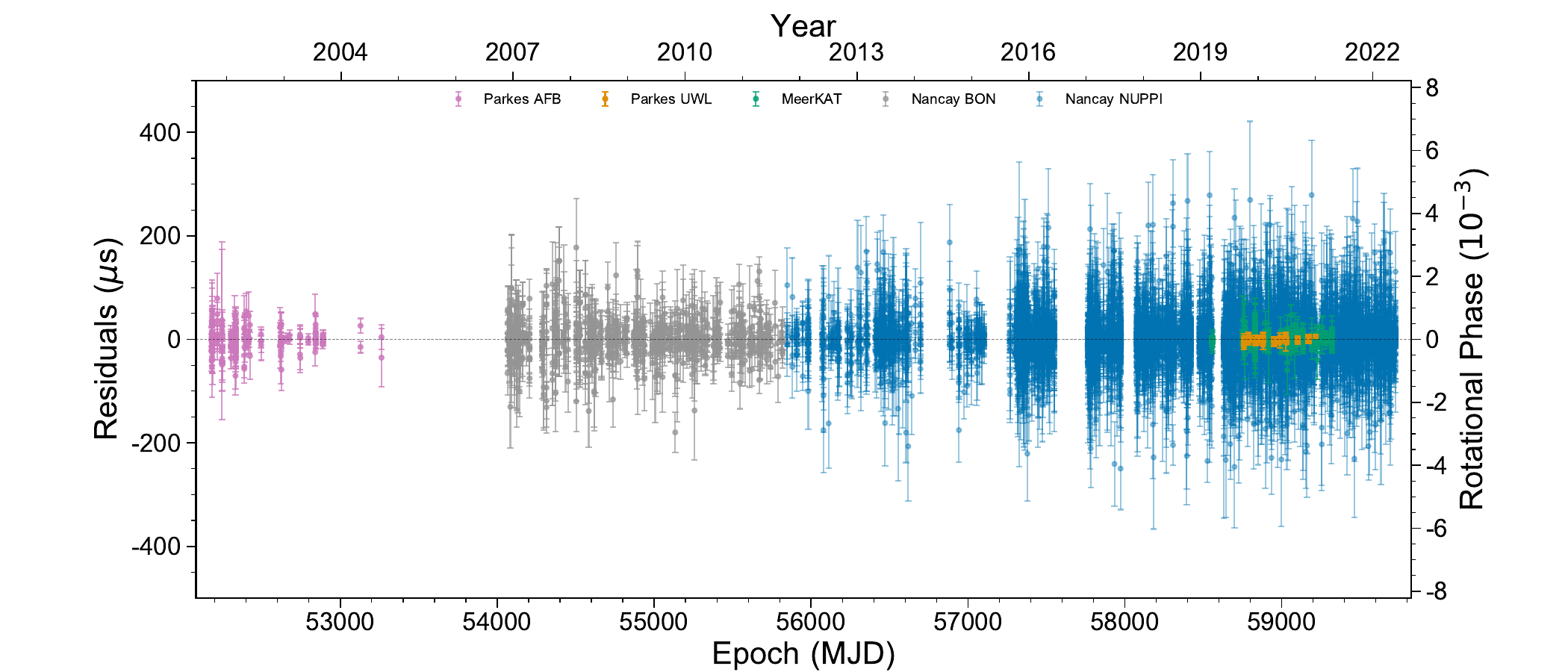}
\includegraphics[width=0.95\textwidth]{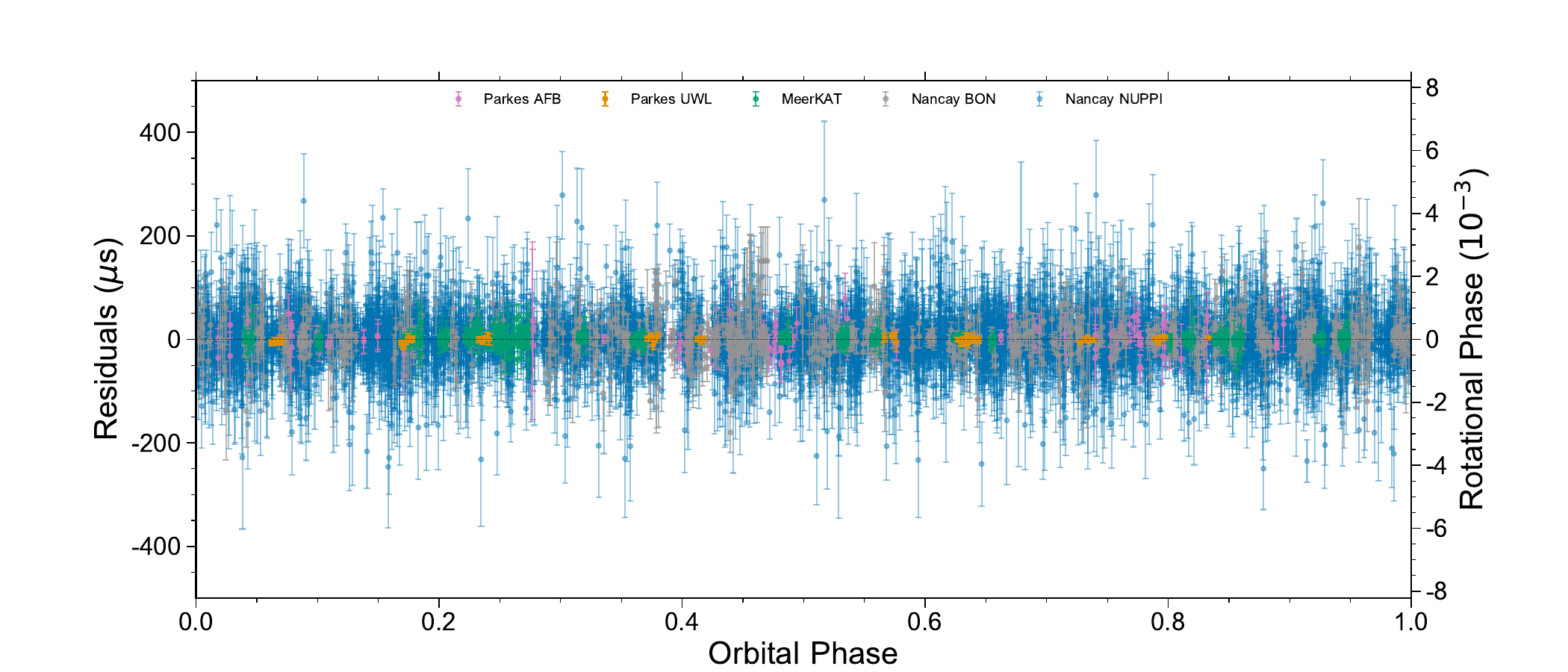}
\caption{Same as Figure~\ref{fig:toas_dates_ophases}, with no time offsets applied to the timing residuals for the different TOA datasets. No trends are visible in the timing residuals plotted as a function of orbital phase or as a function of time, indicating that the timing model describes the data appropriately.}
\label{fig:residus_dates_ophases}
\end{center}
\end{figure*}

\begin{figure}[ht]
\begin{center}
\includegraphics[width=0.95\columnwidth]{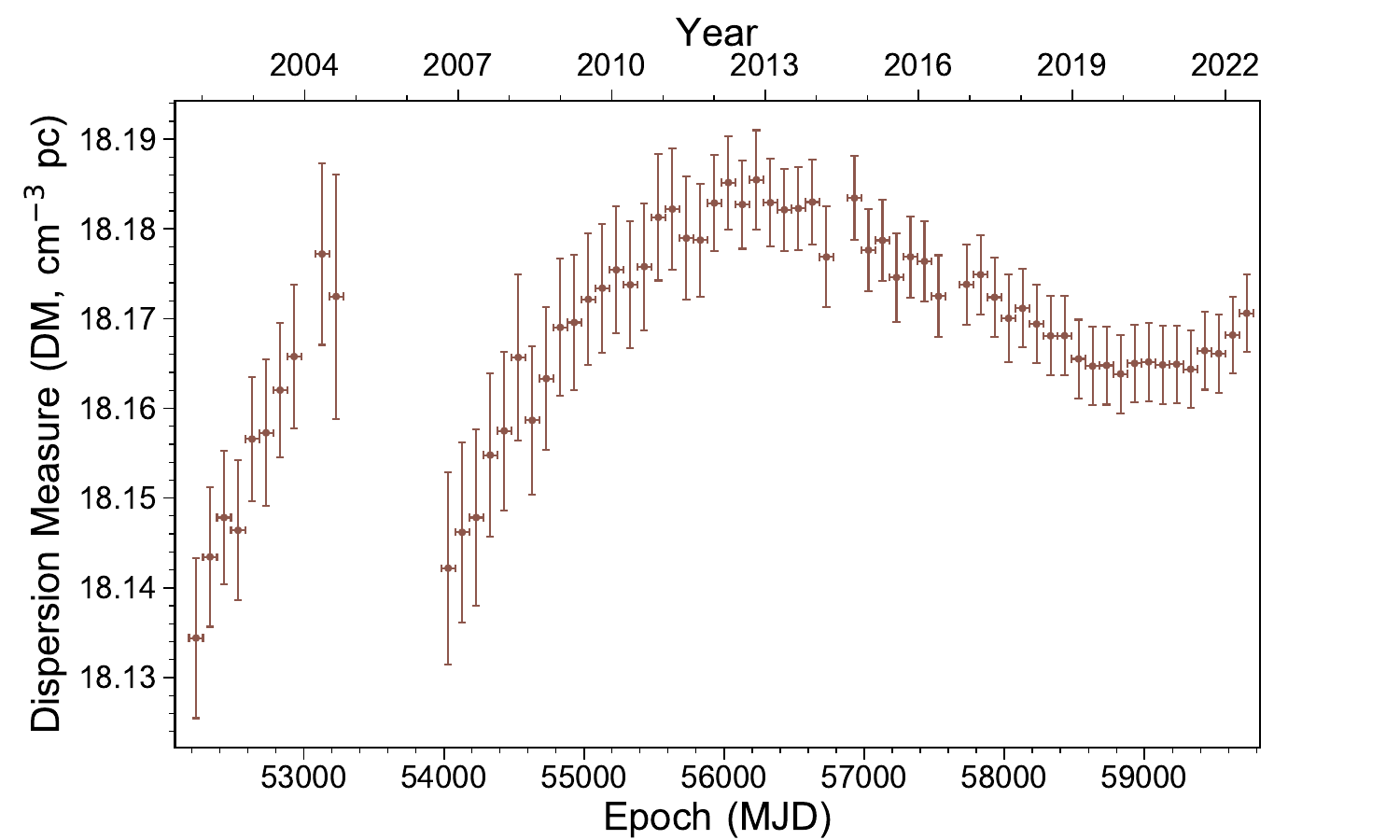}
\caption{Best-fit piecewise-constant DM values from the timing analysis (see Section~\ref{Timing}. We used segments of 100 days, and excluded time intervals that contained no timing data.}
\label{fig:dmx}
\end{center}
\end{figure}

We used \textsc{Tempo2} to fit for the pulsar's sky position and proper motion, its spin parameters (spin frequency, and first and second time derivatives), orbital parameters and PK parameters describing perturbations to the classical Keplerian orbit. The TOAs from the different datasets were aligned with each other using time offsets, that were fitted for as part of the timing analysis. The dispersion measure (DM) and its variations were modeled as a piecewise-constant function, using the ``DMX'' model. The TOA dataset was divided into segments of 100 days, and constant DM values were fitted in each of the intervals containing timing data. In a first iteration of the analysis we obtained a preliminary timing model for \psr that we then used to determine ``EFAC'' factors for each of the individual datasets, \textit{i.e.}, scaling factors applied to the TOA uncertainties such that the individual reduced $\chi^2$ values become equal to unity. The EFAC factors we determined through this procedure are listed in Table~\ref{tab:ToAs}. As can be seen from the table, the EFACs we obtained were all close to 1. A final iteration of the timing analysis with scaled TOA uncertainties yielded the parameters for \psr listed in Table~\ref{tab:ephem}. The differences between measured TOAs and those predicted by the best-fit timing model (the so-called timing residuals) are presented in Figures~\ref{fig:toas_dates_ophases} and \ref{fig:residus_dates_ophases}. The residuals are normally-distributed around 0, demonstrating that the parameters in Table~\ref{tab:ephem} describe the timing data appropriately. Figure~\ref{fig:dmx} shows the best-fit piecewise-consant DM values obtained from the timing analysis. The DM displays long-term variations around the value of 18.163(6) measured by \citet{Jacoby_discovery}.

\begin{table*}
\caption[]{Parameters for \psr from the timing analysis, and derived quantities.}
\label{tab:ephem}
\centering

\footnotesize{
\begin{tabular}{lc}


\hline
\hline
\multicolumn{2}{c}{Observation and data reduction parameters} \\
\hline
Span of timing data (MJD) \dotfill & 52179 -- 59725 \\
Number of TOAs \dotfill & 13478 \\
Reference epoch (MJD) \dotfill & 56000 \\
Time scale \dotfill & TCB \\
Solar system ephemeris \dotfill & DE438 \\
Solar wind electron number density, $n_0$ (cm$^{-3}$) \dotfill & 4 \\
Weighted rms residual ($\mu$s) \dotfill & 9.245 \\
Reduced $\chi^2$ \dotfill & 0.993 \\
\hline
\multicolumn{2}{c}{Astrometric, spin and dispersion parameters} \\
\hline
Right ascension, $\alpha$ (J2000) \dotfill & 15:28:34.95312(6) \\
Declination, $\delta$ (J2000) \dotfill & $-$31:46:06.866(2) \\
Proper motion in $\alpha$, $\mu_\alpha$ (mas yr$^{-1}$) \dotfill & $-$1.39(9) \\
Proper motion in $\delta$, $\mu_\delta$ (mas yr$^{-1}$) \dotfill & $-$3.7(2) \\
Spin frequency, $\nu$ (Hz) \dotfill & 16.4413566796379(11) \\
Spin frequency first derivative, $\dot \nu$ ($10^{-17}$ Hz s$^{-1}$) \dotfill & $-$6.7175(9) \\
Spin frequency second derivative, $\ddot \nu$ ($10^{-28}$ Hz s$^{-2}$) \dotfill & 2.8(6) \\
Dispersion measure, DM (cm$^{-3}$ pc) \dotfill & 18.159\tablefootmark{a} \\
Rotation measure, RM (rad m$^{-2}$) \dotfill & $-$19(1)\tablefootmark{b} \\
\hline
\multicolumn{2}{c}{Binary parameters} \\
\hline
Orbital model  \dotfill & DDH \\
Orbital period, $P_{\rm b}$ (days) \dotfill & 3.1803502(2) \\
Projected semi-major axis of the pulsar orbit, $x$ (lt-s) \dotfill & 11.452301(11) \\
Epoch of periastron, $T_0$ (MJD) \dotfill & 56000.7860(3) \\
Orbital eccentricity, $e$ \dotfill & 0.0002137(2) \\
Longitude of periastron, $\omega$ ($^\circ$) \dotfill & 297.30(4) \\
Rate of advance of periastron, $\dot \omega$ ($^\circ$ yr$^{-1}$) \dotfill & 0.057(3) \\
Rate of change of orbital semi-major axis, $\dot x$ ($10^{-15}$ lt-s s$^{-1}$) \dotfill & $-$2.6(1.9) \\
Orthometric amplitude of the Shapiro delay, $h_3$ ($\mu$s) \dotfill & 1.2(1) \\
Orthometric ratio of the Shapiro delay, $\varsigma$ \dotfill & 0.44(8) \\
\hline
\multicolumn{2}{c}{Derived parameters} \\
\hline
Galactic longitude, $l$ ($^\circ$) \dotfill & 337.9386445 \\
Galactic latitude, $b$ ($^\circ$) \dotfill & 20.2167420 \\
Total transverse proper motion, $\mu_{\rm T}$ (mas yr$^{-1}$) \dotfill & 3.9(2) \\
Position angle of proper motion (J2000), $\Theta_\mu$ ($^\circ$) \dotfill & 200.8(1.8) \\
DM-derived distance (NE2001), $d$ (kpc) \dotfill & 0.8(2)\tablefootmark{c} \\
DM-derived distance (YWM16), $d$ (kpc) \dotfill & 0.8(2)\tablefootmark{c} \\
Mass function, $f$ (M$_\odot$) \dotfill & 0.1594451(5) \\
Spin period, $P$ (ms) \dotfill & 60.822231369658(4) \\
Apparent spin period derivative, $\dot P$ ($10^{-19}$ s s$^{-1}$) \dotfill & 2.4850(3) \\
Total kinematic contribution to spin period derivative, $\Delta \dot P$ ($10^{-21}$ s s$^{-1}$) \dotfill & 1.2(3)\tablefootmark{d} \\
Intrinsic spin period derivative ($\dot P_\mathrm{int}$) ($10^{-19}$ s s$^{-1}$) \dotfill & 2.473(3) \\
Spin-down power, $\dot E$ ($10^{31}$ erg s$^{-1}$) \dotfill & 4.340(5)\tablefootmark{e} \\
Surface magnetic field strength, $B_\mathrm{surf}$ ($10^9$ G) \dotfill & 3.928(2)\tablefootmark{e} \\
Characteristic age, $\tau_c$ ($10^9$ yr) \dotfill & 3.896(5)\tablefootmark{e} \\
Pulsar mass, $M_{\rm p}$ (M$_\odot$) \dotfill & $1.61_{-0.13}^{+0.14}$\ \tablefootmark{f} \\
Companion mass, $M_{\rm c}$ (M$_\odot$) \dotfill & $1.33_{-0.07}^{+0.08}$\ \tablefootmark{f} \\
\hline


\end{tabular}
}

\tablefoot{Numbers in parentheses are the nominal 1$\sigma$ statistical uncertainties on the last quoted digits. 
\tablefoottext{a}{DM variations were modeled using piecewise-constant deviations around this value, as enabled by the ``DMX'' model of \texttt{Tempo2}.}
\tablefoottext{b}{Obtained using the \texttt{RMcalc} program, see Section~\ref{Polarization} for details.}
\tablefoottext{c}{Distances estimated using the NE2001 \citep{NE2001} and YMW16 \citep{YMW16} models, assuming a 20\% uncertainty.}
\tablefoottext{d}{Contributions from the Shklovskii effect \citep{Shklovskii1970} and from the differential Galactic acceleration, as calculated using the \texttt{GalPot} model \citep{McMillan2017,Dehnen1998}. The uncertainty quoted here corresponds to the maximum excursion from the central value, when varying the pulsar distance by $\pm$20\%.}
\tablefoottext{e}{See \citet{handbook} for definitions.}
\tablefoottext{f}{Determined from a Bayesian analysis of the timing data, see Section~\ref{sec:Bayesian} for details.}
}

\end{table*}

The analysis of $\sim$ 16 yrs of timing data, including data of unprecedented quality from the MeerKAT telescope, enabled us to measure the pulsar's proper motion and several PK parameters for the first time. In the following sub-sections we present these measurements in detail and discuss their implications for PSR~J1528$-$3146.

\subsection{Proper motion, distance and spin parameters}
\label{sec:pm}

The timing analysis yields a precise proper motion measurement: we find the total transverse proper motion to be 3.9(2) mas yr$^{-1}$, with a position angle $\Theta_\mu = 200.8(1.8)^\circ$. Given that $0^\circ$ indicates North in the ``observer's" convention we used, the position angle of the proper motion of \psr indicates a Southward motion. Our analysis did not enable us to measure the timing parallax, which otherwise provides a direct constraint on the distance. However, the NE2001 \citep{NE2001} and YMW16 \citep{YMW16} Galactic free electron density models both place the pulsar at a distance $d = 0.8(2)$ kpc, assuming conservative uncertainty levels of 20\%; we use this distance in the
estimates below. 

With these proper motion and distance values, we can estimate the intrinsic spin-down rate of the pulsar, $\dot P_\mathrm{int}$, as:

\begin{equation}
\dfrac{\dot P_\mathrm{int}}{P} = \dfrac{\dot P}{P} - \dfrac{\mu_{\rm T}^2 d}{c} - \dfrac{a}{c} ,
\label{eq:PDOT}
\end{equation}
where $P$ is the rotational period of the pulsar and $\dot P$ is its apparent spin-down rate listed in Table~\ref{tab:ephem}, $\mu_{\rm T}$ is the total transverse proper motion of the pulsar, $d$ its distance, $c$ is the speed of light and $a$ represents the difference of the accelerations of the pulsar and the Solar System in the gravitational field of the Galaxy, projected along the line of sight from the Solar System to the pulsar. 

The term $(\mu_{\rm T}^2 d) / c$ represents the contribution from the Shlovskii effect \citep{Shklovskii1970}, and is relatively small for \psr due to its relatively small distance and transverse proper motion. The last term, $a / c$, can be calculated using the \texttt{GalPot} model of gravitational potential of the Galaxy \citep{McMillan2017,Dehnen1998}, and is also small given the small distance of the pulsar.
The latter two terms combined represent a total kinematic contribution to the spin derivative $\Delta \dot P = 1.2(3) \times 10^{-21}$ s s$^{-1}$ that is small compared to the apparent spin-down rate $\dot P = 2.4850(3) \times 10^{-19}$ s s$^{-1}$, so that our estimate for the pulsar's intrinsic spin period derivative $\dot P_\mathrm{int} = \dot P - \Delta \dot P$ is very close to $\dot P$ (see Table~\ref{tab:ephem}).

The derived intrinsic spin-down rate was then used to determine the pulsar's spin-down power $\dot E$, magnetic field at the stellar surface $B_\mathrm{surf}$, and characteristic age $\tau_c$. Although \psr is located at a small distance to the Earth, its spin-down power $\dot E \sim 4.3 \times 10^{31}$ erg s$^{-1}$ is very low compared to that of known gamma-ray pulsars \citep[see e.g.][]{Fermi2PC}, and no detection of gamma-ray pulsations has been reported to date for this pulsar. The nearest object in the \textit{Fermi} Large Area Telescope (LAT) 12-Year Point Source Catalog \citep[4FGL-DR3, ][]{Fermi4FGL} is located 45' away from the pulsar, meaning that \psr indeed does not have any counterpart in the \textit{Fermi} LAT catalog of gamma-ray sources. 

\subsection{Keplerian and post-keplerian parameters}
\subsubsection{Rate of advance of periastron}

From our timing analysis, we measured $\dot{\omega}_{\rm obs} = 0.057(3)^\circ$~yr$^{-1}$. In the absence of a third object, the observed advance of periastron is given by \cite{handbook}:

\begin{equation}
    \dot{\omega}_{\rm obs} = \dot{\omega}_{\rm GR} + \dot{\omega}_{\rm kin}+ \dot{\omega}_{\rm SO},
    \label{eq:omdot}
\end{equation}
where the first term is caused by the leading order relativistic effect, which, assuming GR, only depends on the total mass of the system, $M_{\rm tot}$ \citep{Robertson1938,Taylor1982}:

\begin{equation}
    M_{\rm tot} = \dfrac{1}{T_\odot} \left(\dfrac{\dot{\omega}_{\rm GR}}{3} (1-e^2)\right)^{3/2} \left( \dfrac{P_{\rm b}}{2\pi} \right)^{5/2} \rm M_{\odot},
\end{equation}
 where $T_\odot={\cal GM}_\odot^{\rm N}/c^3 = 4.925490947... \mu$s is an exact quantity, the solar mass parameter (${\cal GM}_\odot^{\rm N}$, \citealt{Prsa2016}) in time units. Assuming $\dot{\omega}_{\rm GR} \sim \dot{\omega}_{\rm obs}$, we find a total mass $M_{\rm tot}=2.8(2) $ M$_\odot$. This total mass constraint is displayed graphically in Figure~\ref{fig:mpmc} (red lines), and will be compared with the estimated mass from the Bayesian analysis in Section \ref{sec:Bayesian}. 


We now evaluate the above assumption that $\dot{\omega}_{\rm GR} \sim \dot{\omega}_{\rm obs}$. The second term in Equation~\ref{eq:omdot} is caused by kinematic effects and is given by \citep{DDK-1996}:

\begin{equation}
    \dot{\omega}_{\rm kin}=\dfrac{\mu_{\rm T}}{\sin{i}} \cos{(\Theta_\mu-\Omega)},
    \label{eq:omdot_kin}
\end{equation}
%
which is at most $1.2\times 10^{-6}$ $^\circ$ yr$^{-1}$, given the constraints on $i$ and $\Omega$ from the Bayesian analysis presented in Section~\ref{sec:Bayesian}. This is four orders of magnitude smaller than the observed advance of periastron. The last term, $\dot{\omega}_{\rm SO}$, is from spin-orbit coupling. This term is currently only measurable in the case of the double pulsar, PSR~J0737$-$3039A/B \citep{Kramer2021}. Given PSR~J1528$-$3146's much wider orbit compared to those of PSRs~J0737$-$3039A and B, this term about two orders of magnitude smaller. Therefore, the initial $\dot{\omega}_{\rm GR} \sim \dot{\omega}_{\rm obs}$ assumption is warranted.

\subsubsection{Shapiro delay}

The DDH solution uses the orthometric parameterization of \citet{Freire2010} to describe the Shapiro delay. The orthometric amplitude ($h_3$) and the orthometric ratio ($\varsigma$) parameters can be written as a function of the Shapiro range ($r$) and shape ($s$) of the DD model, as follows:

\begin{equation}
h_3 = r\varsigma^3,\\
\varsigma = \dfrac{s}{1+\bar{c}} = \left ( \dfrac{1-\bar{c}}{1+\bar{c}} \right )^{1/2},
\end{equation}
where $\bar{c} \equiv \sqrt{1-s^2}= | \cos{i} |$. From our timing, we obtain significant measurements of both parameters: $h_3 = 1.2(1)$ $\mu$s and $\varsigma = 0.44(8)$. These values can thus be used to constrain the individual masses of the components of the system. In Figure~\ref{fig:mpmc}, these constraints are displayed as blue solid (respectively, dashed) lines for $h_3$ (resp., $\varsigma$).

\subsubsection{Variation of the orbital period}

Our timing analysis of \psr using the DDH model did not yield a significant measurement of the variation of the system's orbital period, $\dot{P}_\mathrm{b}$. The timing model presented in Table~\ref{tab:ephem} therefore does not include a measurement of $\dot{P}_\mathrm{b}$. Fitting for this parameter leads to $\dot{P}_{\rm b,obs}= -0.7(2.2) \times 10 ^{-14}$ s s$^{-1}$, consistent with zero given the present uncertainties. 

The observable orbital period change can be written as follows \citep{handbook}:

\begin{equation}
\left(\dfrac{\dot{P}_{\rm b}}{P_{\rm b}}\right) =  \left(\dfrac{\dot{P}_{\rm b}}{P_{\rm b}}\right)^{\rm GW} +  \left(\dfrac{\dot{P}_{\rm b}}{P_{\rm b}}\right)^{\dot{m}}+\left(\dfrac{\dot{P}_{\rm b}}{P_{\rm b}}\right)^{\rm T} + \dfrac{\mu_{\rm T}^2d+a}{c},
\label{eq:pbdot}
\end{equation}
where the first term corresponds to the intrinsic orbital period variation caused by gravitational wave emission. Assuming the masses for the pulsar and its companion derived from the Bayesian analysis presented in Section~\ref{sec:Bayesian}, we estimate this term to be about $-4.8 \times 10^{-15}\, \rm s \, s^{-1}$ \citep{Peters1964}, an order of magnitude smaller than the uncertainties. 

The second term is the orbital period change caused by mass loss of the system:

\begin{equation}
\left(\dfrac{\dot{P}_{\rm b}}{P_{\rm b}}\right)^{\dot{m}} = \dfrac{8\pi^2G}{T_\odot c^5}\dfrac{I}{M_{\rm tot}}\dfrac{\dot{P}}{P^3},
 \label{eq:pbdot_m}
\end{equation}
where $G$ denotes Newton's gravitational constant, and $I \sim 10^{45}$ g cm$^2$ is the moment of inertia of the pulsar.
Assuming $M_{\rm tot}=2.97$ M$_\odot$ (see Section \ref{sec:Bayesian}), we find that the expected contribution from mass loss to the orbital period change is four orders of magnitude smaller than $\dot{P}_{\rm b,obs}$, and is thus negligible. Similarly, the third term in Equation~\ref{eq:pbdot}, which is due to tidal torques, can be neglected since the companion star is not deformed and is not interacting with the pulsar. 



Finally, the last term corresponds to the kinematic contribution, that includes the Shklosvkii effect and the Galactic differential acceleration. We estimate it to be $\dot{P}_{\rm b, kin} = 5.4(1.4) \times 10^{-15}$ s s$^{-1}$. This is of the same order as $\dot{P}_{\rm b, GW}$, so that the two terms dominate in the calculation of $\dot{P}_{\rm b, obs}$. However, they mostly cancel each other: we find a prediction for $\dot{P}_{\rm b}$ of $\sim 0.7(1.4) \times 10 ^{-15}$ s s$^{-1}$. Our measurement of the apparent orbital period variation is compatible with the predicted $\dot{P}_{\rm b}$, but its uncertainty is still several times larger than that prediction.


%

\subsubsection{Variation of the projected semi-major axis}

The projected semi-major axis $x$ and its secular change $\dot{x}$ have been measured using the DDH model (see Table \ref{tab:ephem}) to be $\dot{x}=2.6(1.9) \times 10 ^{-15}$ lt-s s$^{-1}$. Assuming the absence of a third body in the system, the change in $x$ can be written as the sum of various contributions \citep{handbook}: 

\begin{equation}
    \left(\dfrac{\dot{x}}{x}\right)^{\rm obs} = \left(\dfrac{\dot{x}}{x}\right)^{\rm PM} + \left(\dfrac{\dot{x}}{x}\right)^{\rm GW} + \frac{\mathrm{d}\epsilon_{\rm A}}{\mathrm{d}t} + \dfrac{d\mu_{\rm T}^2+a}{c}+\left(\dfrac{\dot{x}}{x}\right)^{\dot{m}}+\left(\dfrac{\dot{x}}{x}\right)^{\rm SO}.
\label{eq:xdot}
\end{equation}
A detailed evaluation of all these terms was done by \citet{Guo2021} for PSR~J2222$-$0137, whose orbital properties are similar to those of PSR~J1528$-$3146. \citet{Guo2021} concluded that all terms except for the first one are negligible. Similar calculations for \psr enable us to conclude that the same applies to this system. The leading term in Equation~\ref{eq:xdot} describes the secular change in $\dot{x}$ caused by the proper motion of the system, as follows (modified from \citealt{DDK-1996}): 

\begin{equation}
\left(\dfrac{\dot{x}}{x}\right)^{\rm PM} =\mu_{\rm T} \cot{i}\sin{(\Theta_\mu-\Omega)}.
\label{eq:xdotpm}
\end{equation}
We find that $\dot{x}^{\rm PM}$ is at most $3.5 \times 10 ^{-15}$ lt-s s$^{-1}$, comparable to $\dot{x}_{\rm obs}$. This means that the latter measurement already constrains the orbital orientation of the system, in particular the value of $\sin{(\Theta_\mu-\Omega)}$. These constraints are depicted graphically in Figure~\ref{fig:komcosi}.

\subsection{Bayesian analysis}
\label{sec:Bayesian}

\begin{figure*}[ht]
\begin{center}
\includegraphics[width=0.95\textwidth]{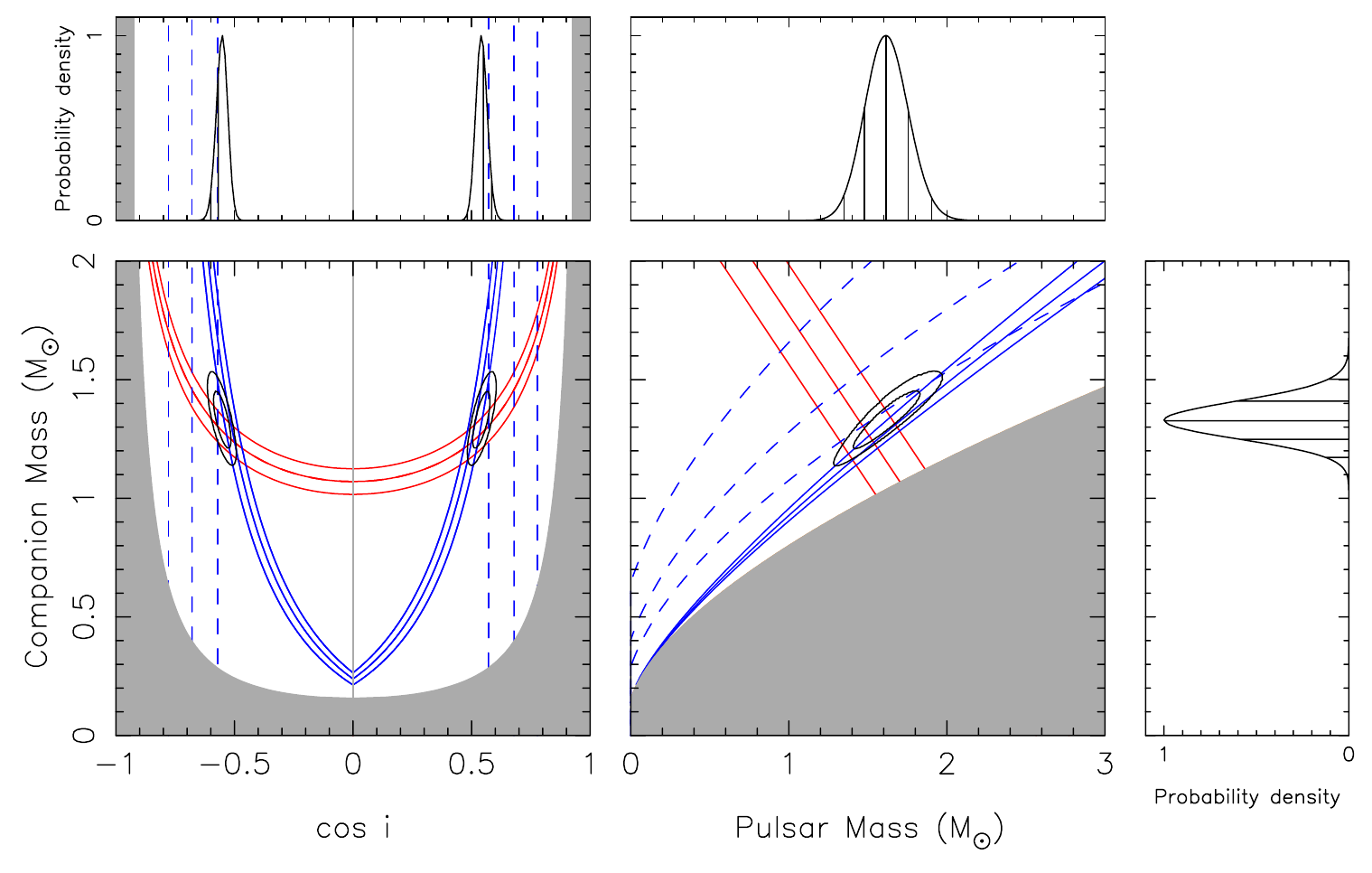}
\caption{Constraints on the mass of the companion of PSR~J1528$-$3146 as a function of the cosine of the orbital inclination (left), and as a function of the mass of the pulsar (right). In the left hand plot, the gray region is excluded by the requirement that the pulsar mass must be greater than zero, and the narrow orange region near $\cos{i} = 0$ is excluded by the marginally-significant constraints on $\dot x$. In the right hand figure, mass values in the gray region are excluded by the mass function measurement. The solid red lines indicate the contours of the 1$\sigma$ uncertainty regions for $\dot \omega$, the 1$\sigma$ uncertainty regions for $h_3$ (resp., $\varsigma$) are displayed as solid blue (resp., dashed blue) lines; these parameters are measured in the DDH orbital model. The solid contours include $1\sigma$ and 2$\sigma$ equivalent percentiles for the joint posterior pdfs for $\cos i$ and $M_\mathrm{c}$ (left) and 
$M_\mathrm{c}$ - $M_\mathrm{c}$ (right). The marginalized posterior pdfs for $\cos i$, $M_\mathrm{c}$ and $M_\mathrm{p}$ are displayed for each plot axis. In the right-hand panel, the probability density increases from right to left.}
\label{fig:mpmc}
\end{center}
\end{figure*}

\begin{figure*}[ht]
\begin{center}
\includegraphics[width=0.7\textwidth]{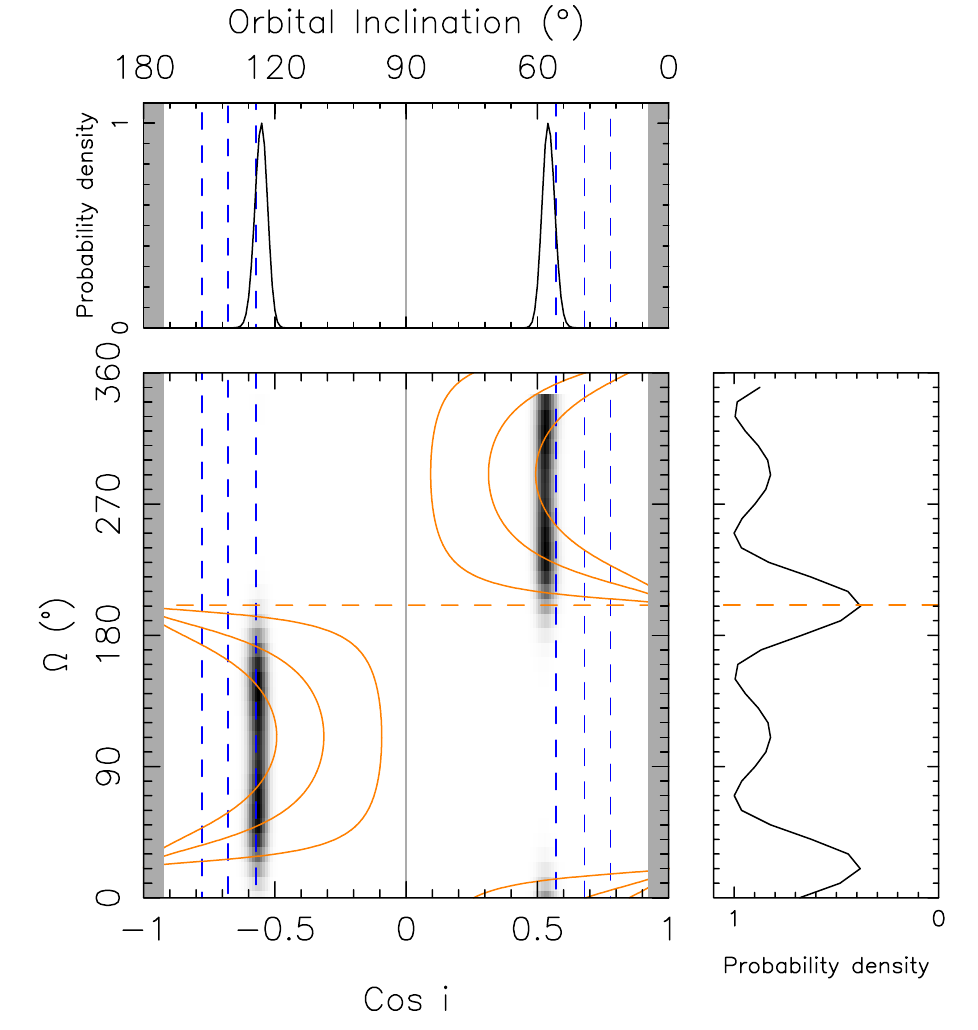}
\caption{Constraints on the longitude of the ascending node, $\Omega$, as a function of the cosine of the orbital inclination. The gray shaded regions are excluded by the mass function and the estimate for the total mass of the binary system. The horizontal dashed orange line indicates the position angle of the proper motion, $\Theta_\mu$. The constraints given by $\dot{x}_{\rm obs}$ and its 1-$\sigma$ uncertainty are displayed as solid orange lines, the constraints given by the measurement of $\varsigma$ are given by the blue dashed lines. These parameters are as measured with the DDH model. The grayscale maps represent the joint posterior pdf for $\cos i$ and $\Omega$, with darker shades corresponding to higher probabilities. The marginal plots display the marginalised posterior pdfs for $\cos{i}$ and $\Omega$. In the right-hand panel, the probability density increases from right to left.}
\label{fig:komcosi}
\end{center}
\end{figure*}


\begin{table*}[]
\centering
\caption{Details of the parameters used in the $\chi^2$ sampling. The $\cos{i}$ and $\Omega$ parameters were sampled over the ranges given in the second and third columns, with respective steps of 0.01 and 10$^\circ$. For the $M_\mathrm{tot}$ parameter we explored values ranging from 2 to 4 M$_\odot$, with a step of 0.01 M$_\odot$. See Section~\ref{sec:Bayesian} for additional details regarding this analysis.}
\begin{tabular}{lcccccc}
\hline
\hline
Region &  $\cos{i}$ & $\Omega$ & Best $\cos{(i)}$ & Best $\Omega$ & Best $M_{\rm tot}$ & $\chi^2_{\rm min}$ \\
\hline
1& $-0.8$ to $-0.4$ &  20$^\circ$ to 110$^\circ$ & $-0.55$ &  70$^\circ$ & 2.98 M$_\odot$ & 13318.23 \\
2& $-0.8$ to $-0.4$ & 110$^\circ$ to 200$^\circ$ & $-0.55$ & 150$^\circ$ & 2.98 M$_\odot$ & 13318.24 \\
3& $0.4$ to $0.8$   & 200$^\circ$ to 290$^\circ$ & $0.55$  & 250$^\circ$ & 2.98 M$_\odot$ & 13318.23 \\
4& $0.4$ to $0.8$   & 290$^\circ$ to 360$^\circ$ & $0.55$  & 330$^\circ$ & 2.98 M$_\odot$ & 13318.24 \\
\hline
\end{tabular}

\label{tab:chi2params}
\end{table*}

In order to estimate the physical parameters of the system (pulsar mass $M_{\rm p}$, companion mass $M_{\rm c}$, orbital inclination angle $i$ and longitude of the ascending node $\Omega$) while self-consistently taking all available constraints into account, we performed a Bayesian analysis of the timing data. Following the method described in \citet{Stovall-2019}, we constructed a $\chi^2$ map of the three-dimensional parameter space of $\cos{i}-\Omega-M_{\rm tot}$. For randomly aligned orbital angular momenta and systems with unknown total mass, these parameters have \textit{a priori} uniform probability densities. These uniform distributions are our Bayesian priors.

The orbital model we used to take the kinematic effects into account was the T2 model implemented in \textsc{Tempo2} \citep{Edwards2006},
which not only takes into account the secular effects on $\dot{x}$ and $\dot{\omega}$ \citep{DDK-1996}, but also effects caused by the varying line of sight to the system caused by the orbital motion of the Earth, such as the annual orbital parallax \citep{DDK-1995}. The latter terms are generally very small and are not included in any other other orbital models. In some cases the annual orbital parallax allows for an elimination of the degeneracy between $i$ and $180^\circ - i$ \citep[see e.g.][]{Stovall-2019}. The T2 model uses $\Omega$ and $i$ as parameters, calculating all kinematic effects internally from these two numbers, which are specified in the ``observer's convention", where $\Omega$ is measured counterclockwise from North through East and $i$ is the angle between the orbital angular momentum and the line of sight from the pulsar to the Earth.

For each value of the $\cos{i}-\Omega-M_{\rm tot}$ grid, we fixed $\Omega$ and $i$ and then derived the PK parameters $M_{\rm c}$, $\gamma$ 
and $\dot{\omega}$ from the total mass, the inclination and the mass function, using the relevant GR equations \citep[from e.g.][]{handbook}. We then fit a solution with these fixed parameters to the timing data, while allowed all the other timing parameters to vary. When \textsc{tempo2} is done minimizing the weighted sum of the squares of all timing residuals, we record the $\chi^2$ value obtained for that point of the $\cos{i}-\Omega-M_{\rm tot}$ grid.

We performed the analysis in two distinct $\cos{i}$ regions instead of a continuous range to reduce the computational cost: from $-0.8$ to $-0.4$ for the first region and from $0.4$ to $0.8$ for the second. We explored total mass $M_{\rm tot}$ values between $2$ and $4$ M$_\odot$. The ranges and step sizes for the three parameters are given in Table \ref{tab:chi2params}.

The resulting three-dimensional $\chi^2$ map was then used to derive the three-dimensional probability density map using the Bayesian likelihood function \citep{Splaver-2002}:

\begin{equation}
  p(\Omega,\cos{i},M) \propto \exp{\left [ (\chi^2_{\rm min}-\chi^2)/2\right]},
\end{equation}
where $\chi^2_{\rm min}$ is the minimum $\chi^2$ value over the entire map. From this three-dimensional joint posterior probability density function (pdf) we derive 2-D joint posterior pdfs for the $M_{\rm c}$ -- $\cos i$ and $ M_{\rm p}$ -- $M_{\rm c}$ planes (see main panels in Figure~\ref{fig:mpmc}), for the $\cos{i}$ -- $\Omega$ plane (see the main panel of Figure~\ref{fig:komcosi}) and marginalized posterior pdfs for $\cos{i}$, $\Omega$, $M_{\rm tot}$, $M_{\rm p}$ and $M_{\rm c}$; most of these appear in the lateral panels of Figs~ \ref{fig:mpmc} and \ref{fig:komcosi}. 

The marginalized posterior pdf of $\cos{i}$ has two maxima, at $-0.55$ and $0.55$. As can be seen from Table~\ref{tab:chi2params}, these solutions have virtually the same $\chi^2$ and therefore the same associated probabilities. The preferred value for $\cos{i}$ of $-0.55$ corresponds to an inclination of $i=123.3_{-4.1}^{+4.8}$$\degree$, the other solution corresponds to $i=56.7_{-4.8}^{+4.1}$$\degree$. The fact that the two maxima have such similar probabilities indicates that the annual orbital parallax is not detected in the considered dataset. As we've seen in section~\ref{Polarization}, the polarimetry suggests that the latter may be the preferred solution.

Fig.~\ref{fig:komcosi} shows four distinct solutions in the $\cos i$ - $\Omega$ space, which are listed in Table \ref{tab:chi2params}. Again, none of the solutions is strongly preferred over the other, as can be seen from Figure~\ref{fig:komcosi}. Furthermore, there is a plateau of high probability between these peaks. The high-probability regions are well described by the $\dot x$ and $\varsigma$ from the timing analysis, the broad distribution of probability can thus be understood to result from the relatively low precision of $\dot{x}_{\rm obs}$.

For the masses we obtain, from the marginalized posterior pdfs, $M_{\rm tot}=2.94^{+0.21}_{-0.20}$ M$_\odot$, $M_{\rm p}=1.61^{+0.14}_{-0.13}$ M$_\odot$ and $M_{\rm c}=1.33^{+0.08}_{-0.07}$ M$_\odot$; where the values are the medians of the pdfs and the uncertainties include 68.3\% of the total probability around these medians. If we include 95.44\% of the probability around the medians, then we obtain $M_{\rm tot}=2.95^{+0.44}_{-0.41}$ M$_\odot$, $M_{\rm p}=1.61_{-0.26}^{+0.28}$ M$_\odot$ and $M_{\rm c}=1.33^{+0.17}_{-0.15}$ M$_\odot$. The value of $M_\mathrm{tot}$ matches that estimated based on the $\dot \omega$ measurement, of $2.8(2)$ M$_\odot$.

\section{Discussion and prospects}
\label{Discussion}

In this work, we have described the polarimetric and timing analysis of our radio observations of PSR~J1528$-$3146. This effort has resulted in much improved physical parameters for this system, and a much improved knowledge of the radio emission properties for the pulsar.

The mass measurements for the \psr system are remarkable for several reasons: they depend mostly on the fact that, even with an orbital eccentricity as low as $2.1 \times 10^{-4}$, we can measure the slow periastron advance of $0.057^\circ \, \rm yr^{-1}$; furthermore, with a low orbital inclination of about $57^\circ$ we were able to measure the Shapiro delay with enough precision to be useful, at least in conjunction with $\dot{\omega}$. Another feature of the system is that the TOA rms of 9.2~$\mu$s represents only 0.00015 of the spin period.
The system is thus reminiscent of PSR~J2222$-$0137 \citep{Boyles2013}, but with less pronounced characteristics: the latter pulsar spins faster ($P = 32.8$ ms), its orbital period is slightly smaller (2.44 days), its orbit is slightly more eccentric ($3.8 \times 10^{-4}$). On the other hand, \psr has a 5 times larger magnetic field. The similarities between the two systems do suggest a close evolutionary relation.

For PSR~J2222$-$0137, the high orbital inclination ($i =  85.27(4)^\circ$), the better timing precision and the slightly more eccentric, faster orbit allow more precise measurements of the Shapiro delay and $\dot{\omega}$. This results in precise mass measurements: $M_{\rm p} = 1.831(10)\, \rm M_{\odot}$ and $M_{\rm c} = 1.319(4)\, \rm M_{\odot}$; the high total mass $M_{\rm tot} =  3.150(14) \rm M_{\odot}$ implies that it is the most massive double degenerate binary known in the Galaxy \citep{Guo2021}. Furthermore, as already highlighted by \citet{Cognard2017}, the amount of material accreted by PSR~J2222$-$0137 is very small, which implies that the mass of that pulsar is basically its birth mass. This shows that NSs are born with a wide range of masses.

For the same reasons, the mass of \psr is also very likely its birth mass - its slower spin and higher magnetic field indicate it has been even less recycled than PSR~J2222$-$0137, and therefore likely accreted less mass. Thus, the precise measurement of the mass of this pulsar will be important, if nothing else it will add to the distribution of known NS birth masses.

The mass of the companion WD is also interesting. First, it could be the most massive WD around any pulsar measured to date. Second,  because, as found by \citet{Mckee-2020}, the masses of WD stars orbiting pulsars do seem to clump around a few preferred values. For He WD companions around MSPs, the measured masses increase with orbital period \citep{TS99}, reaching a maximum of about $0.4\, \rm M_{\odot}$. Then, for several systems with slower spin periods (tens of ms), there is a second clump in companion masses, between 0.7 and $0.9\, \rm M_{\odot}$, and possibly narrower. Finally, PSR~J2222$-$0137 makes a third clump with companion masses around $1.3\, \rm M_{\odot}$. In this respect, the current mass measurement of the WD companion to \psr is already important because it confirms the existence of this third clump, and highlights the general pattern. There are very few WD masses outside these ranges: in two cases, they orbit pulsars that are not recycled, and in very few other cases \citep[like PSR~J1614$-$2230,][]{Crawford2006,Fonseca2018,Shamohammadi2023}, the system is thought to have evolved through case A Roche lobe overflow \citep{Tauris2011}, which is a likely explanation for their very short spin periods.

In the near future, the precision of $\dot{\omega}$ and $\dot{x}$ for this system will keep improving relatively fast, with the uncertainties decreasing faster than the nominal $T^{-3/2}$ rate \citep[where $T$ is the timing baseline, see][]{DT1992}, because of the recent improvement in timing precision achieved by MeerKAT. A better $\dot{\omega}$ will improve the mass measurements, this and a better $\dot{x}$ will improve greatly the constraints on the orbital orientation of the system. The measurement of $\dot{P}_b$ will improve faster, with uncertainties decreasing faster than the nominal rate of $T^{-5/2}$, again because of the timing precision of MeerKAT.

The improved $\dot{P}_b$ that can be achieved in the near future has the potential for a sensitive search for dipolar gravitational wave (DGW) emission, similar to that done with PSR~J2222$-$0137 \citep{Guo2021}. The non-detection of DGW emission in the latter system was important for ruling out the phenomenon of spontaneous scalarisation \citep{Zhao2022}, which was predicted by a class of scalar-tensor gravity theories of \citet{DEF1993}. This happened because of the specific mass of PSR~J2222$-$0137, which is in the so-called ``mass gap'' of spontaneous scalarisation \citep{Shibata2014}, which extends from $1.6$ to $1.9 \rm M_{\odot}$ \citep{Shao2017}. Whether a similar test for \psr is useful or not will depend on the mass of the pulsar, which is not yet known with sufficient precision: the lower half of the uncertainty region is below the mass gap, any value between the median and the 2-$\sigma$ upper limit ($1.89\, \rm M_{\odot}$) would place the pulsar in that gap. Thus, continued timing of this pulsar will be essential in order to improve the mass measurements and determine its potential as a gravitational laboratory.

\begin{acknowledgements}

The MeerKAT telescope is operated by the South African Radio Astronomy Observatory, which is a facility of the National Research Foundation, an agency of the Department of Science and Innovation. SARAO acknowledges the ongoing advice and calibration of GPS systems by the National Metrology Institute of South Africa (NMISA) and the time space reference systems department department of the Paris Observatory. MeerTime data is housed on the OzSTAR supercomputer at Swinburne University of Technology. The Parkes radio telescope is funded by the Commonwealth of Australia for operation as a National Facility managed by CSIRO. We acknowledge the Wiradjuri people as the traditional owners of the Observatory site. This research has made extensive use of NASAs Astrophysics Data System (\url{https://ui.adsabs.harvard.edu/}) and includes archived data obtained through the CSIRO Data Access Portal (\url{http://data.csiro.au}). Parts of this research were conducted by the Australian Research Council Centre of Excellence for Gravitational Wave Discovery (OzGrav), through project number CE170100004 and the Laureate fellowship number FL150100148. The Nan\c{c}ay Radio Observatory is operated by the Paris Observatory, associated with the French Centre National de la Recherche Scientifique (CNRS). We acknowledge financial support from the ``Programme National de Cosmologie et Galaxies'' (PNCG) and ``Programme National Hautes Energies'' (PNHE) of CNRS/INSU, France. The authors acknowledge the constant support from the Max-Planck-Society. MB acknowledges support through ARC grant CE170100004 (OzGrav).

\end{acknowledgements}

\bibliographystyle{aa} 
\bibliography{psr1528.bib}

\end{document}